\documentclass[12pt]{iopart}

\usepackage{graphicx}
\graphicspath{ {images/} }
\begin{document}

\title{Superconductivity from site-selective Ru doping studies in Zr$_5$Ge$_3$ compound}

\author{Sheng Li, Xiaoyuan Liu, Varun Anand, and Bing Lv}

\address{Department of Physics, University of Texas at Dallas, Richardson, Texas 75080, USA}
\ead{blv@utdallas.edu}
\vspace{10pt}
\begin{indented}
\item[]August 2017
\end{indented}

\begin{abstract}
Systematical doping studies have been carried out to search for the possible
superconductivity in the transition metal doped Zr$_5$Ge$_3$ system.  Superconductivity up
to 5.7K is discovered in the Ru-doped Zr$_5$Ge$_{2.5}$Ru$_{0.5}$ sample. Interestingly, with the same
Ru-doping, superconductivity is only induced with doping at the Ge site, but remains absent
down to 1.8K with doping at the Zr site or interstitial site. Both magnetic and transport studies
have revealed the bulk superconductivity nature for Ru-doped Zr$_5$Ge$_{2.5}$Ru$_{0.5}$ sample. The
high upper critical field, enhanced electron correlation, and extremely small electron-phonon
coupling, have indicated possible unconventional superconductivity in this system, which warrants further detailed theoretical and experimental studies.
\end{abstract}

%
%
%
%
%

\section{Introduction}

Searching for new superconducting materials with distinct crystal structures has been a major goal for experimental physicists, and has been proven to be fruitful in achieving superconductivity with higher transition temperatures T$_c$ as seen in cuprate cuprate\cite{HTSMuller} and iron pnictide\cite{Hosono}, or novel physics such as
heavy fermions\cite{heavy} and quasi one dimensional\cite{NbPdS,TaPdTe} superconductors. More superconductors discovered, particularly in a family where large number of compounds exist, will help us to expand the material base of superconductors,  to design new superconductors through comparing the common features in different  superconducting families, and to understand more deeply the high T$_c$ superconductivity mechanism as well. The hexagonal Mn$_5$Si$_3$-type structure, with a large amount of compounds crystallizing in this structure and similar structural derivatives, will be an ideal model system for such searching attempts. In addition, various charge doping possibilities, in both cation/anion sites and interstitial sites in this type of structure have made it to be a rich playground for a variety of physical phenomena such as superconductivity, ferro/antiferromagnetism\cite{ZrPt,Mn5Si3AFM,Mn5Ge3FM}. Among these, Zr$_5$Sb$_3$\cite{Bing}, is represented as the first superconductor discovered in this large family. Chemical doping with Ru in the Sb site in Zr$_5$Sb$_3$, has caused the structural transformation from hexagonal Mn$_5$Si$_3$-type(space group P6$_3$/mcm) to the tetragonal W$_5$Si$_3$-type(space group I4/mcm) and consequently increased the superconducting transition temperature(T$_c$) from 2.3K to 5K\cite{ZrSbRu}. Moreover, several new A$_2$Cr$_3$As$_3$ (A=K, Rb, Cs)\cite{KCrAs}\cite{RbCrAs}\cite{CsCrAs} superconductors with the highest transition temperature, $\sim$ 6.1K, were reported. These new superconductors crystalized in a similar structure,  but without effective bonding between each Cr$_6$As$_6$ chain, thus making them quasi one dimensional superconductors having a very high upper critical field\cite{KCrAs}. The temperature dependence of the change in the penetration depth $\Delta$$\lambda$(T) indicates that there are line nodes in the superconducting gap, implying a possible odd pairing state of these superconductors\cite{huiqui}\cite{huiqui1}. All of these discoveries bring us the confidence that more superconductors will be discovered in this large Mn$_5$Si$_3$-type compound family.

\begin{figure}
\includegraphics[width=6.5in, bb=35 0 500 190]{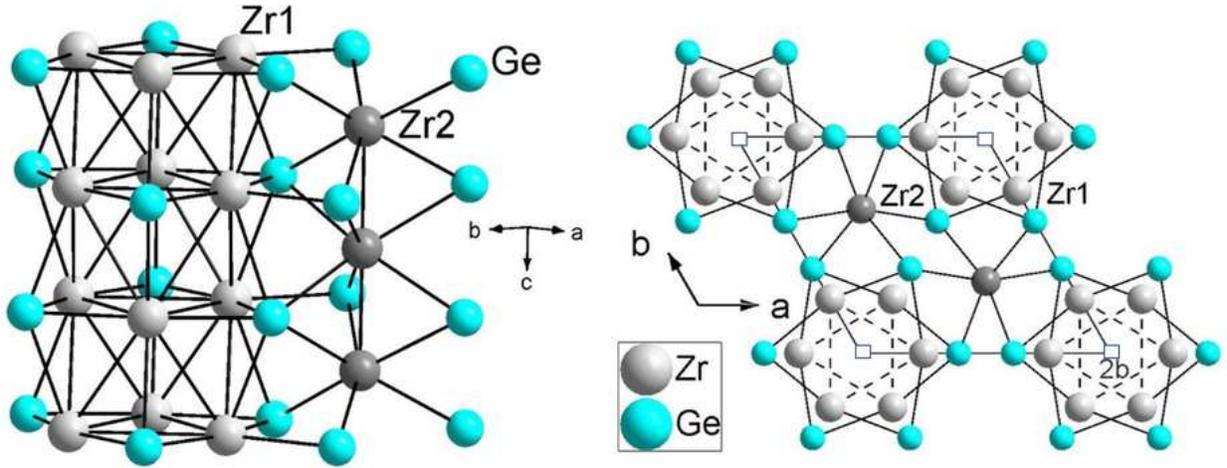}
\caption {(color online) Crystal structure of Zr$_5$Ge$_3$. (Left) side view of Zr$_6$Ge$_6$ and Zr linear chain, (Right) projection of Zr$_5$Ge$_3$ structure. The dotted lines indicate the trigonal antiprism coordination of the Zr$_6$Ge$_6$ chain where an octahedral site forms; small square at the conner of the unit cell show the interstitial 2b site. Two crystallographically different Zr atoms are highlighted as well.} \label{fig1}
\end{figure}

We have chosen Zr$_5$Ge$_3$\cite{ZrGe} as our parent compound to start with for inducing superconductivity. Fig. 1 represents the crystal structure of Zr$_5$Ge$_3$. It features capped trigonal antiprismatic Zr$_6$Ge$_6$ chains along the c-axis, which are subsequently interconnected by another linear chain of Zr.  Unlike A$_2$Cr$_3$As$_3$ where there is no effective bonding between Cr$_6$As$_6$ chains, in the Zr$_5$Ge$_3$, the distance of Zr-Ge between the Zr$_6$Ge$_6$ chains and Zr linear chain is 2.83$\AA$ suggesting weak but effective bonding between Zr$_6$Ge$_6$ and Zr linear chain and a three dimensional nature for Zr$_5$Ge$_3$. Considering the crystal structure described above, there are three ways to change the electronic state at the Fermi surface of Zr$_5$Ge$_3$ through chemical doping. Firstly, directly dope a transition metal at the Zr site. Since Zr is also a transition metal, it should be compatible to do this replacement. Secondly, we can dope transition metals to the Ge site as Zr can form the same intermetallic structure with other transition metal like Zr$_5$Ir$_3$ and Zr$_5$Pt$_3$\cite{ZrPt}. Previous efforts have shown a successful example along this direction with a significant T$_c$ enhancement in the Ru doped Zr$_5$Sb$_{3-x}$Ru$_x$ and Hf$_5$Sb$_{3-x}$Ru$_x$\cite{HfRuSb}system. The last way is to take advantage of the octahedral interstitial sites formed by the Zr$_6$Ge$_6$ trigonal antiprism chains. These unoccupied interstitial 2b sites at the corner of the unit cell, as highlighted in the Fig. 1(b), can host not only small atoms such as B, C, and O, but also bigger transition metal anions such as Bi, Zn, Ru, and Sb\cite{Interstitial}, which further increase the wide electronic tunability in this material. In this paper we present our results of doping transition metal Ru (which is in the middle of Zr and Ge element from period table) at three different sites(Zr, Ge, and interstitial site). We find that superconductivity can only be induced at the Ge site doping, but remains absent down to 1.8K at the Zr site or interstitial site doping with the same Ru-doping level.

\section{Experimental methods and characterization}

The polycrystalline samples were prepared through an arc-melting technique on a water-cooled copper hearth in a home-made arc furnace under argon-atmosphere, with Zr as a getter. Four different samples, which include the parent compound Zr$_5$Ge$_3$,
the Ru doped samples Zr$_{4.5}$Ru$_{0.5}$Ge$_3$, Zr$_5$Ge$_{2.5}$Ru$_{0.5}$, and the interstitial site filled sample Zr$_5$Ge$_3$Ru$_{0.5}$, respectively, were synthesized under the same condition. The starting materials are Zr pieces(99.8$\%$, Strem Chemicals),
Ge pieces(99.99$\%$, Alfa Aesar) and Ru powder(99.9$\%$, Alfa Aesar). These materials were weighed in a glove box filled with argon. During the arc melting process, the melted samples were flipped and remelted several times to ensure the homogeneity of the samples. To avoid the formation of other binary or ternary compounds,
the current was rapidly decreased to zero at the end of process. The total weight loss of the sample was less than 1$\%$ before and after the reaction. The X-ray diffraction measurements were performed on the Rigaku Smartlab,
and the Rietveld refinement of the XRD patterns were done through the JANA 2006 package. The dc magnetic susceptibility $\chi$(T,H) were carried out using the MPMS(Quantum Design) down to 2K and up to 7T. Resistivity as a function of temperature and field $\rho$(T,H), was measured with a PPMS-9T (Quantum Design) down to 1.8K using a standard four-probe technique. The specific heat was measured down to 1.8K and up to 9T based on the DynaCool(Quantum Design).

\section{Results and Discussion}

\begin{figure}
\includegraphics[width=6.5in]{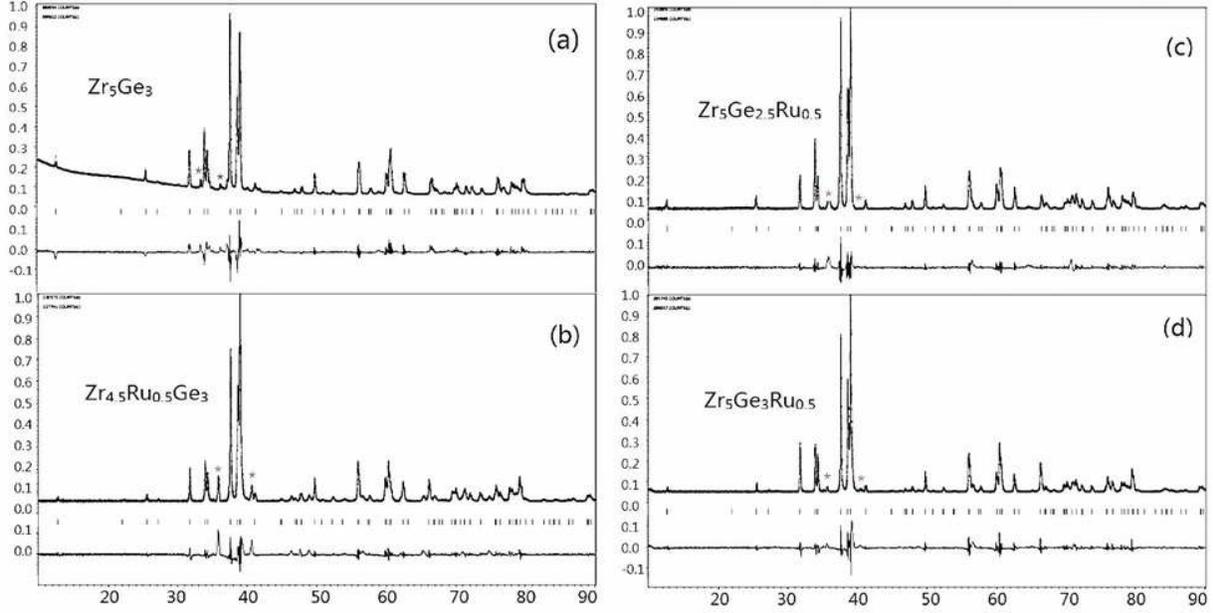}
\caption {(color online) Rietveld refinement of the X-ray diffraction patterns for four samples: (a)Zr$_5$Ge$_3$, (b)Zr$_{4.5}$Ru$_{0.5}$Ge$_3$, (c)Zr$_5$Ge$_{2.5}$Ru$_{0.5}$, and (d)Zr$_5$Ge$_3$Ru$_{0.5}$. Small impurity peaks are marked by blue stars(as ZrGe$_2$) for Zr$_5$Ge$_3$ sample and red stars(as ZrRuGe) for Ru-doped Zr$_5$Ge$_3$ samples, respectively.} \label{fig2}
\end{figure}

\begin{table}
\caption{\label{tabone}Rietveld refinement data of the four samples, small R$_p$ and R$_{wp}$ values indicate high sample quality.}
\begin{indented}
\lineup
\item[]\begin{tabular}{@{}*{5}{l}}
\br
Compound & a($\AA$) & c($\AA$) & R$_p$ & R$_{wp}$ \\
\mr
Zr$_5$Ge$_3$ & 8.0485(9) & 5.6127(6) & 3.71$\%$ & 5.87$\%$ \\
Zr$_{4.5}$Ru$_{0.5}$Ge$_3$ & 8.0832(6) & 5.6511(5) & 4.88$\%$ & 8.20$\%$ \\
Zr$_5$Ge$_{2.5}$Ru$_{0.5}$ & 8.0445(5) & 5.6105(4) & 3.49$\%$ & 6.28$\%$ \\
Zr$_5$Ge$_3$Ru$_{0.5}$ & 8.0556(7) & 5.6234(5) & 5.12$\%$ & 9.75$\%$\\
\br
\end{tabular}
\end{indented}
\end{table}

\begin{figure}
\includegraphics[width=6.5in, bb=10 90 190 225]{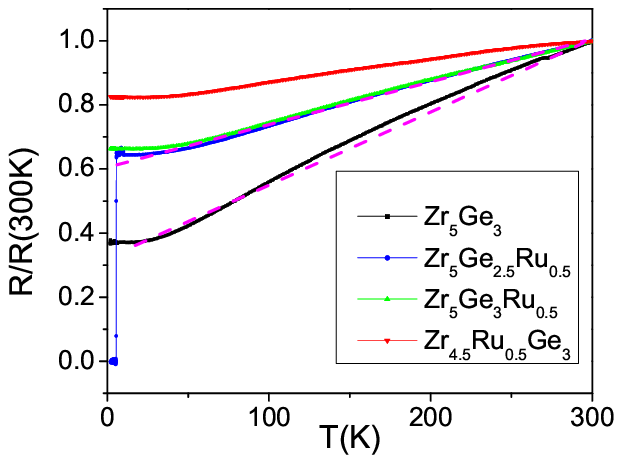}
\caption {(color online) The normalized temperature dependent resistivity data for four different samples with a clear superconducting transition at ~ 5.7K observed in the Zr$_5$Ge$_{2.5}$Ru$_{0.5}$ sample. Two dotted  straight lines were used as the guidance to compare normal state temperature dependence resistivity changes upon doping for parent Zr$_5$Ge$_3$ and doped Zr$_5$Ge$_{2.5}$Ru$_{0.5}$ samples.} \label{fig3}
\end{figure}

The X-ray diffraction patterns and related Reitveld refinements of the four samples are presented in Fig. 2, that shows the synthesized materials adopt the hexagonal Mn$_5$Si$_3$-type structure. The high intensity/sharp peaks of XRD patterns, and the good refinement values of R$_p$ and R$_{wp}$ indicate the high sample quality of the materials.
The refined lattice parameters and R-factors are listed in Table. I.

Small impurities of ZrGe$_2$ (less than 5$\%$) are observed in the parent Zr$_5$Ge$_3$ samples (marked in blue stars in Fig. 2(a)). These impurities are suppressed with Ru doping, but a new impurity phase of ZrRuGe arises in all of the Zr$_{4.5}$Ru$_{0.5}$Ge$_3$, Zr$_5$Ge$_{2.5}$Ru$_{0.5}$, and Zr$_5$Ge$_3$Ru$_{0.5}$ samples. The ZrRuGe impurity is largest in the non-superconducting Zr$_{4.5}$Ru$_{0.5}$Ge$_3$(Fig. 2(b)), but is less than 5$\%$ in both Zr$_5$Ge$_{2.5}$Ru$_{0.5}$(Fig. 2(c)) and Zr$_5$Ge$_3$Ru$_{0.5}$(Fig. 2(d)) samples.
As one can see from Table I, the lattice parameters are increased significantly for Ru doping at the Zr site and octahedral interstitial site as expected, and are decreased slightly when doped at the Ge site, which indicate the successful doping of Ru atoms into the system. SEM-EDX studies have been carried out on these samples and found out the Ru distribution is rather homogenous throughout the samples. The composition is consistent with our nominal 5:3 composition as (Zr+Ru):Ge for Zr$_{4.5}$Ru$_{0.5}$Ge$_3$ sample and Zr:(Ru+Ge) for  Zr$_5$Ge$_{2.5}$Ru$_{0.5}$ sample.

Fig. 3 displays the temperature dependence resistivity data of the four samples from 1.8K to 300K. In order to make a comparison, we normalized the resistivity data to 300K. The resistivity of all the four samples decreases with decreasing temperature as expected for a poor metal. For the impurity phase of ZrGe$_2$,the bulk ZrGe$_2$ is not superconducting, but the sputtered thin film was reported sporadically to be superconducting below 5K\cite{ZrGe2}.
We did not observe the superconducting transition nor resistivity drops in the parent compound, which indicates the ZrGe$_2$ impurity in our Zr$_5$Ge$_3$ sample is not superconducting, and our Zr$_5$Ge$_3$ sample quality is high enough to have a meaningful comparison with doped samples. With the additional Ru doping, the impurity scattering increases which causes the decrease of residual resistance ratio(RRR) value for the Zr$_{4.5}$Ru$_{0.5}$Ge$_3$, Zr$_5$Ge$_{2.5}$Ru$_{0.5}$, and Zr$_5$Ge$_3$Ru$_{0.5}$ samples. Both Zr$_5$Ge$_{2.5}$Ru$_{0.5}$ and Zr$_5$Ge$_3$Ru$_{0.5}$ samples show a almost linear relationship of resistivity with temperature, between 50K to 250K(dotted lines in the Fig. 3), which is different from the parent compound Zr$_5$Ge$_3$, implying the possible enhancement of electron correlation upon doping where superconductivity may arise. Indeed, in the Zr$_5$Ge$_{2.5}$Ru$_{0.5}$ sample, a sudden resistivity drop blow 5.7K to zero resistance is observed, which is a characteristic of superconducting transition. Superconductivity on the other hand, is absent in the Zr$_{4.5}$Ru$_{0.5}$Ge$_3$ and Zr$_5$Ge$_3$Ru$_{0.5}$ samples. Since all the Ru doped samples do have the same impurity phase, and no superconductivity signal is detected in the Zr$_{4.5}$Ru$_{0.5}$Ge$_3$ which contains the highest amount of impurity phase(as seen in Fig. 2(b)), we exclude the possibility that superconductivity is caused by ZrRuGe in the Zr$_5$Ge$_{2.5}$Ru$_{0.5}$ system. Please also be noted that the impurity phase of ZrRuGe was reported to be nonsuperconducting at ambient pressure, but does become a superconductor at about 10K\cite{ZrRuGe} once synthesized under high pressure. It is worthwhile to mention that under the current situation, we could not completely exclude the possibility of interatomic mutual occupancy by Ru at both Zr and Ge sites, or even interstitial sites in our doped samples. Since all of the three doped samples were synthesized under the same condition, the interatomic mutual occupancy would take place in all the samples, if exist. Since we only observed superconductivity in the Ge-site doped sample, it is fair to say that the possible mutual occupancy does not play a significant role, and the Ge-site doping apparently is critical to induce the superconductivity in this system.
\begin{figure}
\includegraphics[width=6.5in, bb=8 110 360 250]{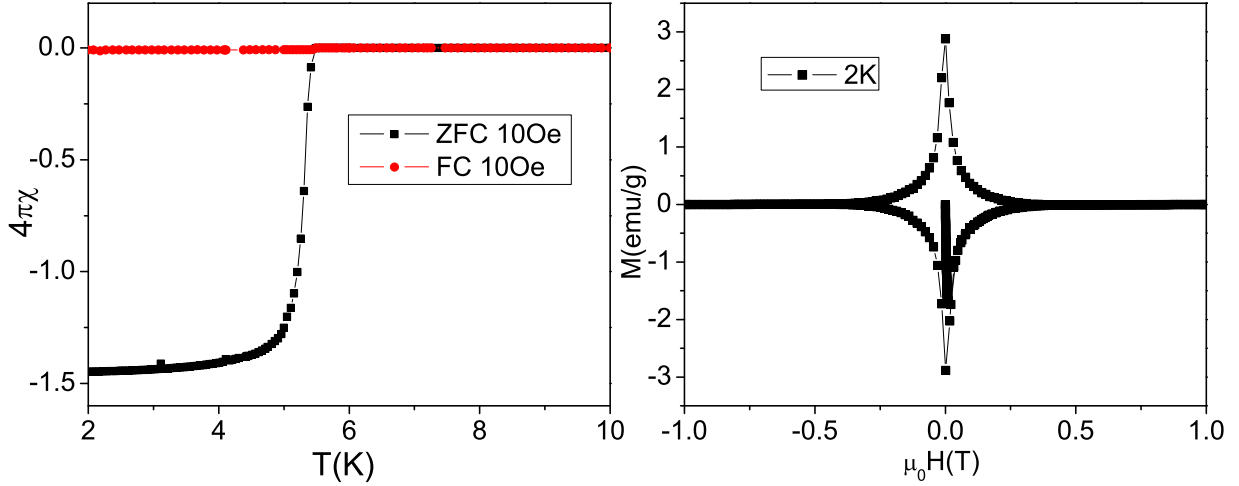}
\caption {(color online) (a) Temperature dependence of the DC
magnetization measured in the ZFC mode and the FC mode at 10Oe. (b) The MH Loop measured with a field
sweeping rate of 50 Oe/s at 2K, which shows a type-II superconductor behavior.} \label{fig4}
\end{figure}

\begin{figure}
\includegraphics[width=6.5in, bb=10 90 210 230]{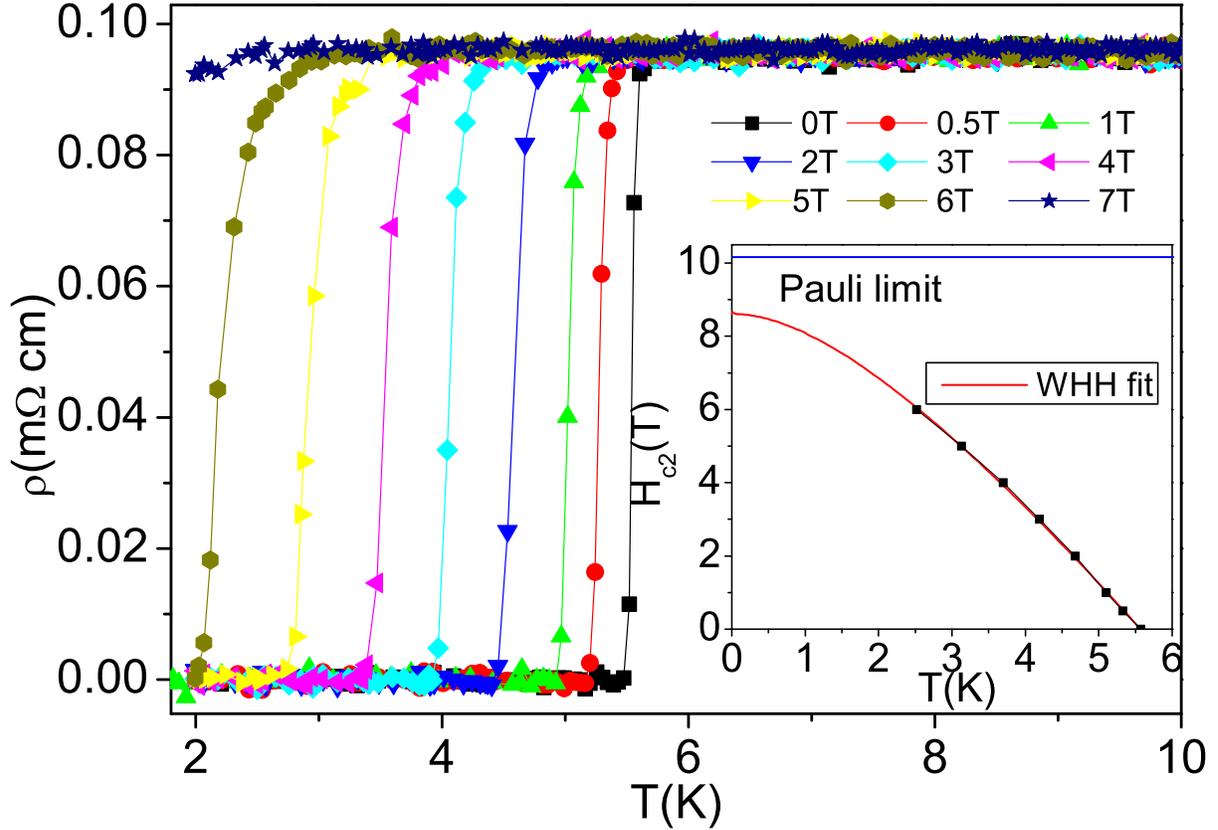}
\caption {(color online) Temperature dependence of resistivity  at different
magnetic fields: 0, 0.5, 1.0, 2.0, 3.0, 4.0, 5.0, 6.0, and 7.0 T.
The insert shows the WHH fitting of the temperature dependence of the critical field with
H$_{c2}$(T) (90$\% \rho_n$), where the straight line shows the Pauli limitation.}
\label{fig5}
\end{figure}

The superconductivity in Zr$_5$Ge$_{2.5}$Ru$_{0.5}$ is further evidenced by the magnetic susceptibility $\chi$(T), electrical resistivity $\rho$(T,H), and the specific heat C$_p$(T) measurements. Fig. 4 displays the $\chi$(T) measured at 10Oe, both in the zero-field-cooled (ZFC) and field-cooled (FC) modes. A clear diamagnetic shift is observed below 5.7 K(Fig. 4(a)). The shielding volume fraction derived from the ZFC curve is about 1.5 before the correction of demagnetization effect at 10Oe, indicating bulk superconductivity of the sample. As the magnetic field increases, we do not observe a dramatic change of transition temperature at low fields. The M-H loop at 2K shows the clear type-II superconductor characteristics(Fig. 4(b)), and the lower critical field H$_{c1}$ is less than 100Oe.

The upper critical field of this newly discovered superconductor is noticeably much higher than the first known Zr-based superconductor with the same  Mn$_5$Si$_3$-type found in Zr$_5$Sb$_3$. However it is lower than the H$_{c2}$ of A$_2$Cr$_3$As$_3$ (A=K, Rb, Cs) superconductors which are more quasi-one dimensional. We further measured the magnetoresistance $\rho$(T,H) to determine the precise
upper critical field. The magneto resistivity data is shown in Fig. 5. One can see that with the highest applied 7T magnetic field, a resistivity jump above 2K is still visible. Taking the 10$\%$ resistivity drop as criteria, we can extrapolate the H$_{c2}$ value
to zero temperature using the Werthamer-Helfand-Hohenberge(WHH) theory. The calculated upper critical field is 8.2T through the WHH fitting, shown as the insert of Fig. 5, which is comparable with the Pauli limit about 10T corresponding to 1.84Tc. One possible reason for such a high upper critical field could be that the Ru doping to the Ge site has weakened the bond between Zr$_2$ atoms and Zr$_6$Ge$_6$ chains, thus making the structure more anisotropic/one dimensional.

\begin{figure}
\includegraphics[width=6.5in, bb=10 90 210 230]{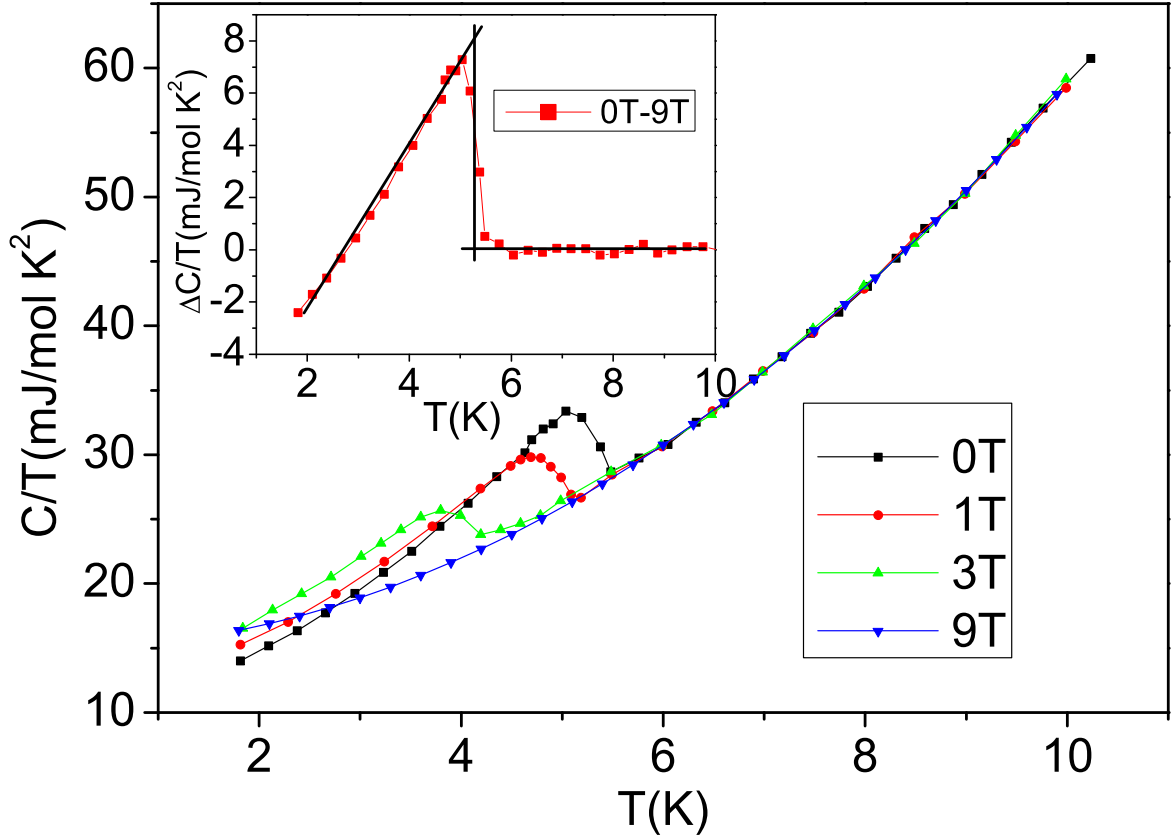}
\caption {(color online) Field dependence of the heat capacity data, the insert shows the electronic specific data.} \label{fig6}
\end{figure}

The bulk superconductivity in Zr$_5$Ge$_{2.5}$Ru$_{0.5}$ can be further demonstrated by the pronounced specific heat anomaly in Fig. 6. Through subtracting the normal state specific
data, we can determine the electronic contributions in the superconducting state. A superconducting specific heat
anomaly appears at 5.4 K with the specific heat jump of about $\Delta C/T_c$ = 8 $mJ/mol K^2$ at zero magnetic field. From the Debye fitting of normal specific
heat data at 9T using $C = \gamma_n T+\beta T^3$, we can get $\gamma_n$ = 14.96 $mJ/mol K^2$,
and $\beta$ = 0.44 $mJ/mol K^4$, which corresponds to the electronic Sommerfeld coefficient
and Debye temperature respectively. The Debye temperature can be deduced from the $\beta$ value through the relationship $\Theta_D$ =
$(12\pi^4k_BN_AZ/5\beta)^{1/3}$, where $N_A$ = 6.02 $\times
10^{23}$ mol$^{-1}$ is the Avogadro constant, and Z is the number of atoms in the molecule. The obtained Debye temperature is about 314K. From $\gamma_n$ and
$\Delta C/T_c$ we get the $\Delta C/\gamma_n T_c$ about 0.53, which is much smaller than the BCS value 1.43. Although the small value of $\Delta C/\gamma_n T_c$ might be underestimated as
the existence of an impurity phase in the material, our XRD pattern and large shielding fraction from magnetic measurements indicate such effect should be small. The small value of $\Delta C/\gamma_n T_c$ reveals that the Zr$_5$Ge$_{2.5}$Ru$_{0.5}$ is a superconductor with a rather weak electron-phonon coupling. It is noteworthy that such a small value of $\Delta C/\gamma_n T_c$ has also been observed in some iron based superconductors such as FeS\cite{FeS} and KFe$_2$As$_2$\cite{KFeAs}, which have similar superconducting transition temperatures and also for the p-wave superconductor Sr$_2$RuO$_4$\cite{SrRuO}.

In order to investigate why the superconductivity happens, we also
measured the specific heat of Zr$_5$Ge$_3$. The specific heat data of Zr$_5$Ge$_3$ is shown together with the normal state specific heat of Zr$_5$Ge$_{2.5}$Ru$_{0.5}$ in Fig. 7.
A broad hump around 4K of Zr$_5$Ge$_3$ specific heat data may be caused by the Schottky anomaly. One can clearly see the enhancement of the Sommerfeld coefficient in the Zr$_5$Ge$_{2.5}$Ru$_{0.5}$ sample.
A reasonable fitting of Zr$_5$Ge$_3$ specific heat data gives us $\gamma_n$ = 12.85 $mJ/mol K^2$,
$\beta$ = 0.31 $mJ/mol K^4$, respectively. The Debye temperature calculated from $\beta$ is about 353K.
Apparently the Sommerfeld coefficient increases but the Debye temperature decreases upon the Ru-doping into the Zr$_5$Ge$_3$ system. The decreased Debye temperature indicates the decreasing of the electron-phonon coupling which is consistent with the small $\Delta C/\gamma_n T_c$ value we obtained from specific heat analysis. On the other hand, as the Sommerfeld coefficient is proportional to
the density of state(DOS)at the Fermi level, the  15$\%$ enhancement of $\gamma_n$ upon Ru doping implies the enhancement of the DOS, which may be responsible for the observed superconductivity in Zr$_5$Ge$_{2.5}$Ru$_{0.5}$ sample.

\begin{figure}
\includegraphics[width=6.5in, bb=10 90 190 225]{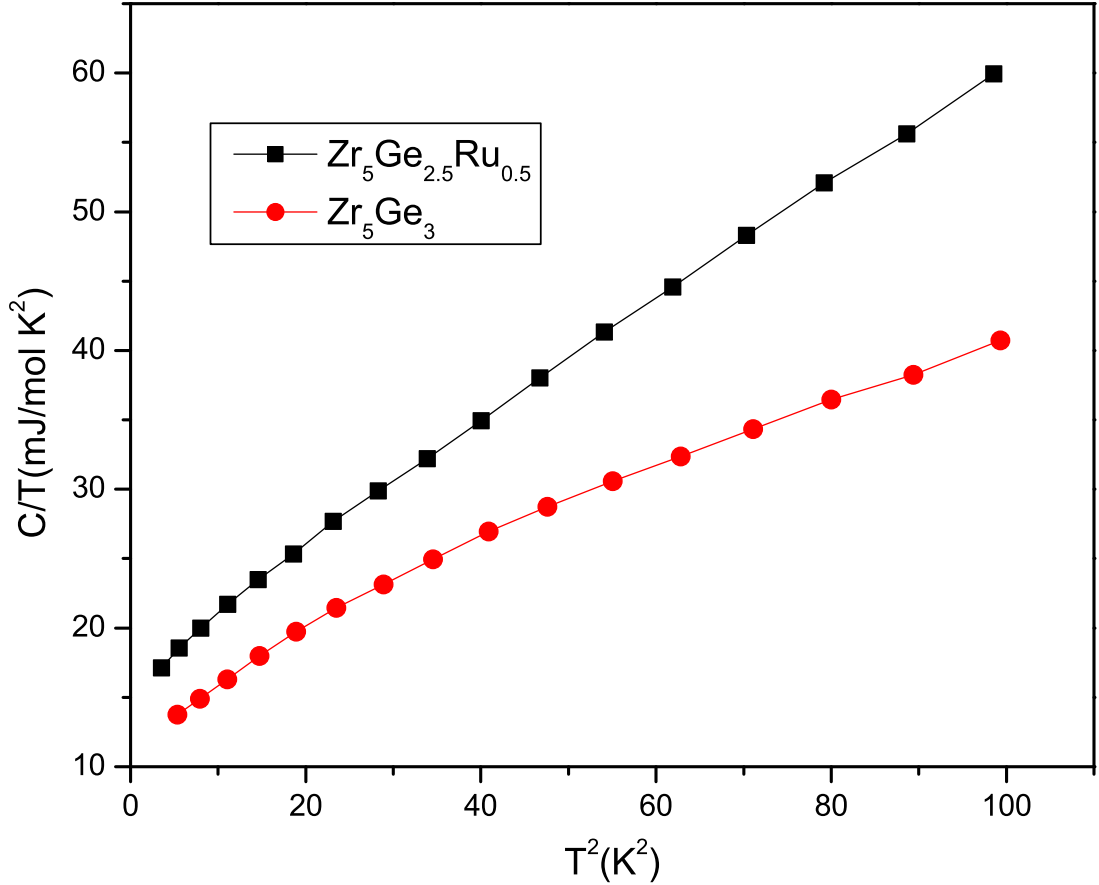}
\caption {(color online) The specific heat data of Zr$_5$Ge$_3$ at zero field, and the normal state specific data of Zr$_5$Ge$_{2.5}$Ru$_{0.5}$ at 9T. } \label{fig7}
\end{figure}

Since the Ru-doping at either Zr site or interstitial site(nonsuperconducting sample) is electron doping, and Ru-doping at the Ge site(superconducting sample) brings more holes to the system, one could imagine that the Fermi level of Zr$_5$Ge$_3$ is located on a negative slope in the density of states. Hole doping causes the shift of the Fermi level to the DOS peak, which is consistent with the observed increasing DOS from specific heat analysis. However, the weak electron-phonon coupling might hinder the further increase of the T$_c$ following the BCS picture. One can take the other methods, such as doping to the Zr site with rare earth metals, to test whether the superconductivity can be induced through other hole-doping approaches. In that way, we will be able to find out how critical the Ru element is for the induced superconductivity in these system as suggested by others\cite{HfRuSb}. Alternatively, as one can see from Table I, only Ru doping at a Ge site can cause a slight shrink of the lattice parameter, chemical pressure may also plays  a role here for the induced superconductivity. Therefore, it will be worthwhile to investigate external pressure effects on both the superconducting  and nonsuperconducting samples, to see whether the T$_c$ of the superconducting sample can be further enhanced, and whether the superconductivity could be induced in the nonsuperconducting samples under high pressure or not.

In summary, we have synthesized Zr$_5$Ge$_3$ and doped it with Ru at different sites. Interestingly, superconductivity can only be introduced through Ru doping to the Ge site with T$_c$ up to 5.7K. Magnetization, electrical resistivity, and specific heat measurements have confirmed the bulk superconductivity in the sample. The enhancement of the DOS at the Fermi level through hole doping of Ru may be responsible for the induced superconductivity. The nearly linear correlation of resistivity with the temperature at high temperature, the high upper critical field, and weak electron-phonon coupling, all indicate possible unconventional superconductivity occurred in the Zr$_5$Ge$_{2.5}$Ru$_{0.5}$ samples which warrants further theoretical and experimental investigation of this system.

\section{Acknowledgments}
The authors would like to thank D. Z. Wang and Z. F. Ren for the help with specific heat measurement, and R. Glosser for critical reading of the mauniscript. This work is supported by US Air Force Office of Scientific Research Grant No. FA9550-15-1-0236 and the start-up funds from University of Texas at Dallas.

\section{References.\label{bibby}}

\end{document}